\documentclass[lettersize,conference]{IEEEtran}
\usepackage{amsmath,amsfonts}
\usepackage{algorithmic}
\usepackage{algorithm}
\usepackage{array}
\usepackage[caption=false,font=small,labelfont=sf,textfont=sf]{subfig}
\usepackage{textcomp}
\usepackage{stfloats}
\usepackage{url}
\usepackage{verbatim}
\usepackage{graphicx}
\usepackage{cite}
\usepackage{multirow}

\def\BibTeX{{\rm B\kern-.05em{\sc i\kern-.025em b}\kern-.08em
    T\kern-.1667em\lower.7ex\hbox{E}\kern-.125emX}}
    
\usepackage{tikz}
\usepackage{array}

\begin{document}

% \title{Carbon-Efficient Integration and Aging-Aware Replacement of Processing-in-DRAM Accelerators in Transformer Inference Servers}

% \title{Scaling Carbon Analysis of Processing-in-DRAM Accelerators Across GPU Baselines and Transformer Workloads}

\title{Sustainable Processing‑in‑DRAM Integration and Replacement for Transformer Inference Servers}

\author{\IEEEauthorblockN{Bipin Thapa Magar}
\IEEEauthorblockA{\textit{Department of Electrical and Computer Engineering} \\
\textit{University of Kentucky},
Lexington, KY, USA \\
bipin.thapa@uky.edu}
\and
\IEEEauthorblockN{Ishan Thakkar}
\IEEEauthorblockA{\textit{Department of Electrical and Computer Engineering} \\
\textit{University of Kentucky},
Lexington, KY, USA \\
igthakkar@uky.edu}
}

\maketitle

\begin{abstract}

Processing-in-DRAM (PID) accelerators are emerging as a sustainable alternative to GPUs for transformer model inference. They bring compute to data, mitigate the memory bandwidth limitation, and reduce the overall energy footprint. Yet, carbon analysis of PID accelerators is covered for only a limited set of hardware and workloads. In this work, we present a carbon modeling framework for PID-enabled inference servers that captures embodied carbon along with idle and active operational carbon. Our evaluation spans three GPU baselines, six PID architectures, and eight transformer workloads. Additionally, we propose an aging-aware model that finds the carbon-optimal replacement period for PID accelerators over a 10-year horizon. Across evaluated configurations, our analysis shows that including PIDs in AI inference servers increases the total embodied carbon by at most 2.5\%, while the total carbon is reduced by up to 1.14$\times$ for encoder and 1.04$\times$ for decoder transformer workloads. The results also depend on GPU selection for pairing, where introducing PID modules in high-end HGX-H100 servers incurs 5.4--5.9$\times$ higher total carbon compared to A100 and H100 servers, due to their larger idle and embodied carbon. Introducing PID modules in higher-end GPU servers, compared to the A100-only baseline, increases (decreases) the total carbon-delay product (CDP) for encoder workloads (decoder workloads).
%For encoder workloads, the carbon-delay product (CDP) increases with higher-end GPUs because GPUs remain idle as PIDs perform the encoder computation. For decoder workloads, PIDs reduce the total CDP by at least 2.6$\times$ compared to the A100-only baseline.
The aging-induced carbon-optimal PID replacement varies greatly with PID design and workload. Therefore, the sustainable deployment of PIDs requires a co-design of PID architecture, GPU pairing, and replacement policy.
\end{abstract}


\section{Introduction}

\IEEEPARstart{T}{ransformer} models have revolutionized the field of artificial intelligence (AI) by achieving state-of-the-art performance across workloads, including text-only large language models, vision transformers, and multimodal encoders \cite{tu2024overview,chatterji2025chatgpt}. As these models are actively deployed in data centers, the requirements go beyond computation alone. Recent reports estimate that data centers consume around 4.4\% of total electricity in the U.S., which is forecasted to reach 6.7--12\% by 2028 \cite{iea2024,shehabi2024}. As data centers grow, global CO$_2$ emissions from them are estimated to reach 720 million tons by 2030 \cite{liu2020energy}. Hence, minimizing carbon emissions from AI inference in data centers is urgent for sustainability \cite{li2025ai}.

% \IEEEPARstart{A}{rtificial} Intelligence (AI) inference workloads have diversified rapidly, moving beyond text-only large language models to encompass vision transformers and multimodal encoders deployed at data center scale \cite{tu2024overview,chatterji2025chatgpt}. This growth is not merely computational. Recent industry and government analyses estimate that data centers already draw roughly 4.4\% of U.S. electricity, with that share projected to reach 6.7 to 12\% by 2028 as AI deployments continue to scale \cite{iea2024,shehabi2024}, pushing global data center emissions toward 720 million tons of CO$_2$ by 2030 \cite{liu2020energy}. Reducing the energy and resource cost of AI inference has therefore become an urgent priority for sustainable computing and data center design \cite{li2025ai}.

Current AI inference largely runs on graphics processing unit (GPU) based servers. Although GPUs have growing compute density, they are limited by the memory wall \cite{huang2019gpu}, especially during the memory-bound autoregressive decoding phase of transformer inference \cite{pope2023efficiently}. Processing-in-DRAM (PID) accelerators address this problem by placing compute logic inside or near the DRAM array \cite{atria,transpim,primate,neupims,attacc,attenpim,star,artemis,magar2026dristi,sting}. This results in lower data movement, which is a dominant factor in energy consumption during memory-bound execution. Prior PID works have proposed designs for both the compute-bound encoder \cite{transpim,artemis,primate} and the memory-bound decoder \cite{neupims,attacc,attenpim} phase of transformer inference.

% Much of today's AI inference runs on graphics processing unit (GPU) centric servers, but GPU throughput is increasingly bottlenecked by the memory wall \cite{huang2019gpu}, particularly during the memory-bound autoregressive decoding phase of transformer inference \cite{pope2023efficiently}. Processing-in-DRAM (PID) accelerators respond to this bottleneck by embedding compute logic directly within or near the DRAM array, cutting the data movement that dominates energy in memory-bound execution \cite{atria,transpim,primate,neupims,attacc,attenpim,star,artemis,magar2026dristi,sting}. A growing body of work has proposed PID designs specialized for either the encoder \cite{transpim,artemis,primate} or decoder \cite{neupims,attacc,attenpim} phase of transformer inference, each targeting a different point in the compute-bound and memory-bound design space.

However, few studies have explored how deploying such PID accelerators in cloud-to-edge settings impacts sustainability and carbon efficiency. Current carbon modeling mainly focuses on CPU- and GPU-based systems \cite{gupta2022act,eeckhout2022first,ji2024scarif,li2025ecoserve}, with only a few focusing on edge PID applications\cite{ollivier2022sustainable,patel2025lifecycle}. Recent work has extended this modeling to PID-integrated servers \cite{thapamagar2026carbon}, but it limits the study to a single GPU and a single workload, limiting the generalization of its findings. This leaves an open problem: whether the carbon benefits of PID integration hold across multiple GPUs and varying workloads with different parameter counts and modalities. 
% This work aims to answer this by evaluating PID integration across a larger set of hardware and workloads than prior work, for both cloud and edge deployments.

% However, the sustainability implications of deploying these accelerators as part of full inference servers, particularly in cloud-to-edge settings, remain poorly understood. Architectural carbon modeling has matured considerably for CPU and GPU-based systems \cite{gupta2022act,eeckhout2022first,ji2024scarif,li2025ecoserve} and for edge PID platforms \cite{ollivier2022sustainable,patel2025lifecycle}, and recent work has begun to extend this modeling to PID integrated servers \cite{thapamagar2026carbon}. These studies, though informative, are typically anchored to a single GPU generation and a single representative workload per server class. This leaves open a central question: \emph{Do the sustainability and carbon benefits of PID integration observed for one GPU and one workload persist as GPU baselines scale and as the workload mix broadens across parameter counts and modalities?} Answering this question requires evaluating PID integration across a substantially larger hardware and workload space than has been considered to date, and doing so in a way that is relevant to both cloud and edge deployments.

To address this gap, this paper models embodied and operational carbon for six PID designs and three GPU baselines, where the GPUs span single-GPU and multi-GPU server configurations. These PID--GPU server configurations are evaluated against eight transformer workloads, covering both encoder and decoder classes with language and vision modalities. We also perform aging-aware PID replacement analysis that evaluates the carbon-optimal PID replacement interval for the various PID architectures, GPU baselines, and workloads over a ten-year deployment horizon. To enable our analysis to reflect realistic upgrades rather than hypothetical designs, we added PID modules to GPU inference servers as drop-in installations of compute-enabled DRAM devices.

% To address this gap, this paper models embodied and operational carbon for six PID designs, three encoder-oriented and three decoder-oriented, integrated with three GPU baselines that span single GPU and multi-GPU server configurations. The resulting servers execute eight transformer workloads that cover encoder and decoder classes and both language and vision modalities. On top of this carbon model, we further extend an aging-aware replacement analysis to characterize how the carbon-optimal PID replacement interval shifts with PID architecture, GPU baseline, and workload class over a ten-year deployment horizon. Because PID modules are realized as compute-logic-enhanced DRAM devices, they can be added to GPU inference servers via drop-in installation in spare DRAM slots, enabling our analysis to reflect realistic upgrade paths rather than hypothetical clean-slate designs.

Our contributions are as follows:
\begin{itemize}
    \item We present a unified carbon modeling framework for PID-enabled transformer inference servers that spans six PID designs, three GPU baselines, and eight representative encoder and decoder workloads.
    \item We quantify how PID integration changes total carbon and carbon-delay product relative to GPU-only servers, and show that GPU baseline choice dominates absolute emissions even when PID reduces operational carbon.
    \item We evaluate the carbon-optimal replacement period for aging-aware analysis of PID subsystems and show how the optimal period depends on co-design of PID choices, GPU baselines, and workloads.
    % \item We extend aging-aware analysis to PID subsystems and show that the carbon-optimal PID replacement interval is strongly design and workload dependent, underscoring that sustainable PID deployment requires co-design of PID architecture, GPU pairing, and replacement policy.
\end{itemize}

\section{Background}

\subsection{Carbon Modeling for Computing Systems}
The carbon footprint of a system can be modeled in terms of two components: embodied and operational carbon. Embodied carbon captures the carbon emissions from the extraction of raw materials, device manufacturing, and transport, whereas operational carbon captures the carbon cost of operating the device. As technology advances, the carbon cost of fabrication is rising, causing embodied carbon to rival or even exceed operational carbon \cite{gupta2021chasing,bardon2020dtco}. This makes it necessary to look beyond energy efficiency toward overall carbon efficiency.

% A system's carbon footprint has two components. Embodied carbon captures emissions from raw material extraction, fabrication, and transport of hardware, while operational carbon captures emissions from electricity used during operation. As process nodes shrink and fabrication becomes more resource intensive, embodied carbon has grown to rival or even exceed operational carbon \cite{gupta2021chasing,bardon2020dtco}, so energy efficiency alone is insufficient for carbon efficiency.

Existing architectural carbon models estimate embodied carbon from device properties, including die area and process node, and combine them with fabrication carbon intensity and industry data. Operational carbon, on the other hand, is derived from application energy and the deployment grid's carbon intensity \cite{gupta2022act}. These models have been further extended by introducing fixed-time and fixed-work assumptions \cite{eeckhout2022first}, modeling 3D and 2.5D integration \cite{zhao20243d}, and including accelerator subsystems \cite{ji2024scarif,li2025ecoserve}.

% Architectural carbon models estimate embodied carbon from die area, process node, fabrication carbon intensity, and industry data, and combine it with operational carbon derived from application energy and the deployment grid's carbon intensity \cite{gupta2022act}. Later work formalized the embodied and operational tradeoff under fixed work and fixed time assumptions \cite{eeckhout2022first}, extended embodied modeling to 3D and 2.5D integration \cite{zhao20243d}, and added accelerator subsystems at server scale \cite{ji2024scarif,li2025ecoserve}. In parallel, GPU vendors have begun releasing product carbon footprint summaries, such as NVIDIA's report for HGX H100 \cite{nvidia2025hgxh100pcf}, providing direct embodied carbon data for accelerator platforms and further grounding architectural models in measured product-level footprints.

The operational carbon of AI inference servers is evaluated based on application energy, and these servers rarely run at peak utilization. During idle periods, the servers consume idle power, which is often comparable to active power for low-utilization workloads. Recent studies have therefore included idle carbon as a first-order term \cite{shehabi2024,thapamagar2026carbon}, while others prioritize extending hardware lifetime over early replacement to amortize embodied carbon \cite{ji2024towards}. \textit{This study evaluates the effects of idle carbon and analyzes how to optimize the replacement period of the low-embodied-carbon PID devices}.

% Inference servers rarely run at peak utilization, and idle power can be comparable to active power. Recent studies therefore treat idle carbon as a first-order term \cite{shehabi2024,thapamagar2026carbon} and advocate amortizing embodied carbon by extending hardware service life rather than following a fixed replacement schedule \cite{ji2024towards}. Our aging-aware analysis in Section~\ref{sec:aging_results} takes a complementary view, treating the PID replacement interval as a design variable enabled by its low embodied overhead.

The performance drop of devices such as processors and DRAM as they age has been extensively studied \cite{masrur2012schedulability,fieback2017dram}. This aging-induced degradation has been shown to increase operational carbon in servers, as application time grows with degrading performance \cite{ollivier2022sustainable,patel2025lifecycle,thapamagar2026carbon}. This analysis has also been extended to PID inference servers in both edge and cloud configurations, but these studies are largely limited to fixed hardware and workload configurations \cite{ollivier2022sustainable,patel2025lifecycle,thapamagar2026carbon}. This motivates examining the generalizability of the results across broader hardware and workload scenarios.

% Building on multi-year aging data for CPUs, GPUs, and DRAM \cite{masrur2012schedulability,fieback2017dram}, recent work folds aging-driven degradation into server-level carbon models, where components require increasing active time as they age, raising operational carbon over long deployments \cite{patel2025lifecycle,thapamagar2026carbon}. Existing carbon evaluations of processing-in-memory hardware are largely limited to edge devices or fixed server settings \cite{ollivier2022sustainable,patel2025lifecycle,thapamagar2026carbon}, motivating the broader hardware and workload study presented in Section~\ref{sec:experimentation}.

\subsection{Processing-in-DRAM for Transformer Inference}
Transformer inference has two phases with distinct hardware demands: encoder (prefill) and decoder (autoregressive). The encoder phase processes the entire input sequence at once as general matrix-matrix multiplication (GEMM) and is mainly compute-bound. GPUs efficiently handle these GEMM operations unless the sequence length is long enough to require multiple off-chip transfers \cite{pope2023efficiently}. The decoder phase, by contrast, generates one output token at a time and requires multiple weight and KV-cache fetches, making it memory-bound \cite{pope2023efficiently}. PID accelerators reduce this data movement by embedding compute logic within or near the DRAM array \cite{atria,transpim,primate,neupims,attacc,attenpim,star,artemis,magar2026dristi,sting}.

% Transformer inference has two phases with distinct hardware demands. The encoder, or prefill, phase processes an entire input sequence at once and is mainly compute-bound, dominated by general matrix-matrix multiplication (GEMM) operations that GPUs handle efficiently until long sequences force repeated off-chip transfers \cite{pope2023efficiently}. The autoregressive decoding phase instead generates output one token at a time and is memory-bound, repeatedly fetching weights and KV-cache data for comparatively little arithmetic work \cite{pope2023efficiently}. PID accelerators target this imbalance by embedding compute logic within the DRAM array or buffer die, reducing energy-intensive off-chip movement \cite{atria,transpim,primate,neupims,attacc,attenpim,star,artemis,magar2026dristi,sting}.

Encoder-oriented PID designs \cite{transpim,primate,atria} compute the prefill phase directly inside DRAM. TransPIM \cite{transpim} implements a token-based dataflow on high-bandwidth memory (HBM) by using bank-level units that handle reduction and softmax. PRIMATE \cite{primate} builds on TransPIM's\cite{transpim} dataflow and adds in-memory dynamic token pruning that discards low-relevance tokens by using HBM2-based top-$k$ engines. ATRIA \cite{atria} is a bit-parallel stochastic accelerator for convolutional neural networks, later adapted in \cite{artemis} to support transformer encoders.

% Encoder-oriented PID designs accelerate the prefill phase directly in memory. TransPIM \cite{transpim} implements a token-based dataflow on high-bandwidth memory (HBM) with bank-level units handling reduction and softmax. PRIMATE \cite{primate} builds on this dataflow with in-memory dynamic token pruning, using HBM2-based top-$k$ engines to discard low-relevance tokens. ATRIA \cite{atria} was originally a bit-parallel stochastic accelerator for convolutional neural networks; here, we adapt its substrate to support transformer encoders.

Decoder-oriented PID designs focus on the memory-bound decoding phase. NeuPIMs \cite{neupims} introduces co-execution between a neural processing unit (NPU) and a PID, offloading attention to the PID subsystem while the NPU handles GEMM layers. AttAcc \cite{attacc} implements general matrix-vector multiplication (GEMV) engines across the HBM hierarchy and performs softmax on the buffer die. AttenPIM \cite{attenpim} introduces configurable bank-level units to exploit multi-level parallelism.

% Decoder-oriented designs focus on the memory-bound decoding phase. NeuPIMs \cite{neupims} enables joint neural processing unit (NPU) and PID execution, offloading attention to the PID subsystem while the NPU handles GEMM layers. AttAcc \cite{attacc} instantiates general matrix-vector multiplication (GEMV) engines across the HBM hierarchy and computes softmax on the buffer die. AttenPIM \cite{attenpim} introduces configurable bank-level units that exploit both token- and head-level parallelism.

These PID architectures differ in terms of DRAM technology, compute logic, and GPU involvement. As these factors jointly affect the carbon footprint, the carbon profile of a PID-enabled server depends on the joint configuration of PID design, GPU, and workload. This motivates the evaluation space in Section \ref{sec:experimentation}.

% These PID architectures differ in DRAM technology, added logic, and GPU involvement during inference. Because each factor independently affects server emissions, the carbon profile of a PID integrated server depends jointly on PID design, GPU baseline, and workload mix, motivating the broad evaluation space in Section~\ref{sec:experimentation}.

\section{Carbon Modeling Framework} \label{sec:methodology}

\subsection{Carbon Modeling}
Our carbon model separates manufacturing and use-phase emissions and then recombines them into a single quantity for each server configuration. The total carbon footprint combines the embodied carbon footprint ($CF_{em}$), which accounts for carbon emitted during manufacturing, and the operational carbon footprint ($CF_{op}$), which accounts for carbon emitted during operation. Their sum gives the total carbon footprint, shown in Equation~\ref{eq:total_carbon}.
\begin{equation}
    CF_{total} = CF_{em} + CF_{op}
    \label{eq:total_carbon}
\end{equation}

Embodied carbon values for the CPU, GPU, and other platform components are drawn from published reports and prior work~\cite{li2025ecoserve,gupta2022act,nvidia2025hgxh100pcf}, with relevant details given in Section~\ref{sec:experimentation}. For PID designs, we additionally model the embodied carbon contributed by the PID compute-logic itself, based on its chip area overhead ($\Delta A_\text{pid}$), as shown in Equation~\ref{eq:pid_emb}. The total PID die area is the sum of the DRAM area ($A_{DRAM}$) and the PID compute-logic overhead ($\Delta A_\text{pid}$), scaled by the number of stacks or chips ($N_{chips}$), as shown in Equation~\ref{eq:pid_area}. The carbon footprint per unit area ($CF_{memory}^{area}$) for each DRAM technology is likewise drawn from prior work and industry-reported data~\cite{li2025ecoserve,gupta2022act,park2022192,samsung_lpddr5}.
\begin{align}
  CF^{pid}_{em} &= Area_{pid} \times CF_{memory}^{area}
  \label{eq:pid_emb} \\
  Area_{pid} &= N_{chips} \times \left( A_{DRAM} + \Delta A_{pid} \right)
  \label{eq:pid_area}
\end{align}

Operational carbon captures how long each component is active or idle and how much power it draws in each state over the deployment period. A server rarely runs at a single steady power draw. Each component (CPU, GPU, PID) alternates between active and idle periods as it serves batches of requests over deployment period $T$~\cite{thapamagar2026carbon}. Rather than modeling a full power trace, we follow~\cite{thapamagar2026carbon} and represent each component $c$ with two states: \emph{active}, where it draws workload-dependent power ($P_{act}^{c}$) for duration $t_{act}^{c}$, and \emph{idle}, where it draws idle power ($P_{idle}^{c}$) for duration $t_{idle}^{c}$. Operational carbon is then a weighted combination of the two, as shown in Equation~\ref{eq:op_carbon}.
\begin{equation}
    CF_{op} = \left[\sum_{c \in \{cpu,gpu,pid\}} P_{act}^{c} \cdot t_{act}^{c} + P_{idle}^{c} \cdot t_{idle}^{c}\right] \cdot CI_{use}
    \label{eq:op_carbon}
\end{equation}
Here $t_{act}^{c}$ and $t_{idle}^{c}$ are the active and idle durations of component $c$, and $CI_{use}$ is the carbon intensity of the electricity grid at the deployment site.

Utilization provides a compact way to express active and idle time. Active time ($t_{act}^{c}$) is set by a utilization factor $U \in [0,1]$, reflecting how often the server serves requests over $T$, as shown in Equation~\ref{eq:active_time}. Idle time ($t_{idle}^{c}$) is the remaining fraction of the deployment period, as shown in Equation~\ref{eq:idle_time}.

\begin{equation}
    t_{act}^{c} = U \cdot T
    \label{eq:active_time}
\end{equation}
\begin{equation}
    t_{idle}^{c} = T - t_{act}^{c} = (1-U) \cdot T
    \label{eq:idle_time}
\end{equation}

In real inference servers, idle carbon can be as high as, or even higher than, the total active contribution for high idle-power components such as the GPU. This is because the servers are actively computing for only around 40\% of the total runtime \cite{shehabi2024}.

% Idle carbon can rival or exceed the active contribution for components such as the GPU that draw substantial power even when not computing. This is particularly significant for inference servers, which are typically active only around 40\% of the time~\cite{shehabi2024}, and whose idle power draw varies across CPU and GPU generations.

To compare architectures on both sustainability and performance, we also track a joint carbon–latency metric. Beyond total carbon, evaluating carbon-efficient architectures requires accounting for performance. Carbon-Delay Product (CDP) jointly captures carbon emission and execution time for a given system~\cite{gupta2022act}, and is calculated as shown in Equation~\ref{eq:cdp}.
\begin{equation}
    CDP = CF \cdot t_{exec}
    \label{eq:cdp}
\end{equation}
Here $CF$ is the carbon footprint of a configuration and $t_{exec}$ is the execution time of the workload on that configuration.

\subsection{Server Configurations: Encoder and Decoder Topologies}
\begin{figure}[t]
    \centering
    \includegraphics[width=\columnwidth]{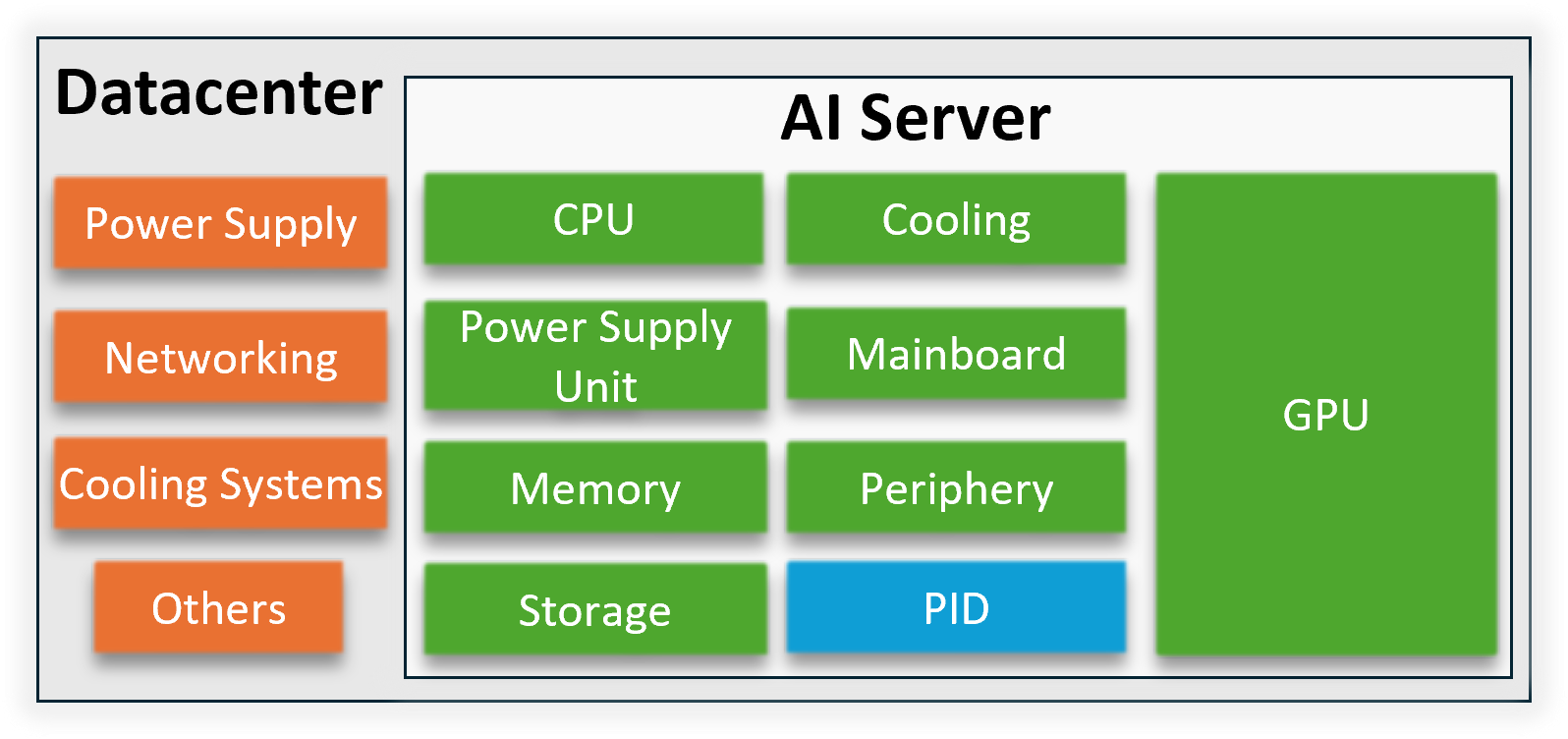}
    \caption{Components of an AI server within a datacenter. Green blocks indicate subsystems modeled for carbon (CPU, GPU, memory, storage, cooling), blue highlights the added PID accelerator, and orange indicates datacenter infrastructure excluded from the analysis.}
    \label{fig:server}
\end{figure}

We now describe how PID and GPU accelerators are integrated into servers for encoder and decoder workloads. Fig.~\ref{fig:server} summarizes a typical datacenter and the scope of our analysis. Each AI server includes the green components in Fig.~\ref{fig:server}, and we add a PID accelerator (blue) that enables efficient transformer inference alongside the GPU. In our setting, PID subsystems are compute-logic-modified DRAM modules that are installed in the memory slots of standard CPU/GPU servers, providing an upgrade solution without requiring a full system redesign. Datacenter-level infrastructure highlighted in orange is excluded from our carbon model.

% We now describe how PID and GPU accelerators are integrated into servers for encoder and decoder workloads. Fig.~\ref{fig:server} summarizes a typical datacenter and the scope of our analysis. Each AI server includes the green components in Fig.~\ref{fig:server}, and we add a PID accelerator (blue) that enables efficient transformer inference alongside the GPU. In our setting, PID subsystems are compute-logic-enhanced DRAM modules installed in additional memory slots of standard GPU servers, providing a drop-in upgrade path without full platform redesign and aligning with existing datacenter deployment and replacement practices. Datacenter-level infrastructure highlighted in orange is excluded from our carbon model.

Each server executes one of two classes of transformer workloads: encoder or decoder. GPUs are well suited to the compute-bound encoder phase, but as sequence length increases, attention activations exceed on-chip memory capacity, forcing repeated transfers to and from off-chip DRAM. Several PID accelerators~\cite{transpim,primate,atria} have therefore been designed to accelerate encoder workloads directly in memory. In our configurations, encoder workloads are fully executed by the PID accelerator and the GPU remains idle throughout.

Decoder workloads, by contrast, are memory bound at each autoregressive step. PID accelerators are well suited to this regime, but decoding still requires an initial encoder, or prefill, pass where GPU execution is beneficial. Accordingly, the GPU and PID accelerator \cite{neupims,attacc,attenpim} co-execute the decoder workload in sequence. The GPU completes the compute-bound prefill pass, after which the PID accelerator performs memory-bound autoregressive decoding. The two components are never active at the same time; co-execution here denotes this division of labor across phases rather than concurrent operation.

The CPU plays the same role in encoder and decoder configurations. It does not run transformer computations, but instead handles orchestration, tokenization, and data scheduling between the GPU and PID subsystem. We therefore model the CPU as active with $U = 0.06$ in all configurations.

\subsection{Aging Model} \label{sec:aging_model}
Finally, we incorporate component aging into the carbon model to study PID replacement policies over multi-year deployments. We extend \cite{thapamagar2026carbon} and model component aging as a year-over-year increase in active time, as shown in Equation~\ref{eq:aging_active_time}.
\begin{equation}
    t^{y}_{act} = t^{1}_{act} \times (1+\alpha)^{y-1}
    \label{eq:aging_active_time}
\end{equation}
Here $t^{1}_{act}$ is the first-year active time, $y$ indexes the deployment year, and $\alpha$ is a component-specific degradation rate. Following prior characterization \cite{fieback2017dram,masrur2012schedulability}, we set $\alpha = 0.06$ for CPU and GPU \cite{masrur2012schedulability} and $\alpha = 0.115$ for DRAM and PID \cite{fieback2017dram}, applied uniformly across workloads.

This aging model induces a time-varying operational footprint over the deployment horizon. The resulting operational carbon over a $T$ year horizon is given in Equation~\ref{eq:aging_operational}.
\begin{equation}
    CF_{op}^{aging} = \sum_{y=1}^{T} \left[\sum_{c \in \{cpu,gpu,pid\}} P_{act}^{c} \cdot t_{act}^{y,c} + P_{idle}^{c} \cdot t_{idle}^{y,c}\right] \cdot CI_{use}
    \label{eq:aging_operational}
\end{equation}

We assume CPU and GPU lifetimes span the full deployment horizon and are never replaced, so their embodied carbon is amortized over $T$. This is because the large embodied carbon of the CPU and GPU far outweighs the operational benefits of replacing them. In contrast, the PID subsystem accounts for approximately 2.5\% of the server's total embodied carbon, which makes its replacement a viable strategy. The PID subsystem is replaced at an interval of $LT_{pid}$, which resets its aging but incurs an additional embodied cost, as given by Equation~\ref{eq:aging_total}.

% We assume CPU and GPU lifetimes span the full deployment horizon and are never replaced, so their embodied carbon is amortized over $T$. The embodied carbon of the CPU and GPU is substantial, and replacing either would incur costs that far outweigh any operational benefit in our setting. The PID subsystem, by contrast, accounts for only a small fraction of the server's total embodied carbon (approximately 2.5\%), making periodic replacement a viable strategy. The PID subsystem may therefore be replaced at interval $LT_{pid}$. Each replacement resets its aging but incurs an additional embodied cost, as given by Equation~\ref{eq:aging_total}.
\begin{equation}
    CF^{aging}_{total} = CF^{emb}_{cpu} + CF^{emb}_{gpu} + \frac{T}{LT_{pid}} \cdot CF^{emb}_{pid} + CF_{op}^{aging}
    \label{eq:aging_total}
\end{equation}

The PID replacement interval $LT_{pid}$ thus directly impacts the PID embodied cost and the operational carbon because each replacement resets aging. We analyze this tradeoff between embodied and operational carbon in Section~\ref{sec:aging_results} to identify the carbon-optimal PID replacement interval across PID--GPU--workload combinations.

% The replacement interval $LT_{pid}$ thus governs a direct tradeoff between aging-driven operational carbon and recurring embodied cost. We analyze this tradeoff in Section~\ref{sec:aging_results} to identify the carbon-optimal PID replacement interval for each PID–GPU–workload combination.

\section{Evaluation and Results} \label{sec:experimentation}
\subsection{Evaluation Setup}
\textbf{Baselines:} 
We evaluate three GPU baselines: NVIDIA A100 SXM4~\cite{nvidia_a100}, NVIDIA H100 SXM5~\cite{nvidia2022h100whitepaper}, and NVIDIA HGX-H100 8$\times$~\cite{nvidia2025hgxh100pcf}, each paired with an Intel Xeon Max 9480 CPU~\cite{intel_xeon_max_9480} and 220~GB of main memory as the fixed host configuration. Against these baselines, we evaluate six PID accelerators: three encoder-oriented designs (ATRIA~\cite{atria}, TransPIM~\cite{transpim}, and PRIMATE~\cite{primate}) and three decoder-oriented designs (NeuPIMs~\cite{neupims}, AttAcc~\cite{attacc}, and AttenPIM~\cite{attenpim}). Each PID design is configured according to the specifications reported in its original publication, so that differences in our results reflect architectural choices made by the original authors rather than adjustments we introduced. However, we modify ATRIA~\cite{atria} since it was originally designed for convolutional neural networks, and we scale it to a 64 GB configuration with an added softmax unit for supporting transformer inference. For NeuPIMs\cite{neupims}, we substitute the GPU baseline for the NPU used in its original design.

\textbf{Workloads:}
We evaluate eight transformer workloads: four encoder and four decoder. The workload set is chosen to span multiple modalities (language and vision) and a range of parameter counts within each class. Encoder workloads include BERT, RoBERTa-PubMed, ViT-Base, and SigLIP-So400M, while decoder workloads include GPT-3-7B, Gemma-2-9B, Mistral-7B, and LLaMA-3-8B. All workloads use a batch size of 16 and are executed the same number of times for a fair comparison across architectures. Table~\ref{tab:workloads} summarizes the configurations for each of the eight transformer models.

\begin{table}[t]
    \centering
    \caption{Configurations of the Transformer workloads evaluated in this study. N: Sequence length.}
    \label{tab:workloads}
    \begin{tabular}{|c|c|c|c|c|c|c|}
        \hline
        \textbf{Workload} & \textbf{Heads} & \textbf{Layers} & \textbf{$d_{model}$} & \textbf{$d_{ff}$} & \textbf{N} \\ \hline
        BERT              & 12 & 12 & 768  & 3072  & 128  \\ \hline
        RoBERTa-PubMed    & 12 & 12 & 768  & 3072  & 3072 \\ \hline
        ViT-Base          & 12 & 12 & 768  & 3072  & 197  \\ \hline
        SigLIP-So400M     & 16 & 27 & 1152 & 4304  & 729  \\ \hline
        GPT-3-7B          & 32 & 32 & 4096 & 16384 & 128  \\ \hline
        Gemma-2-9B        & 16 & 42 & 3584 & 14336 & 8192 \\ \hline
        Mistral-7B        & 32 & 32 & 4096 & 14336 & 4096 \\ \hline
        LLaMA-3-8B       & 32 & 32 & 4096 & 14336 & 8192 \\ \hline
    \end{tabular}
\end{table}

\textbf{Simulation Setup:}
We build a custom Python simulator for each server configuration. Each of the eight workloads is run through every applicable PID--GPU pairing, with the same number of executions fixed per workload class. The simulator tracks each data transfer, memory access, multiply-accumulate, and non-linear operation, and outputs the total latency and dynamic power for each workload. These values may differ from each PID's original implementation due to modeling variations, but these variations are applied uniformly across architectures, making the comparison valid. Future work can explore silicon-validated modeling. The output is then fed into the carbon model described in Section~\ref{sec:methodology} to compute the resulting carbon footprint. For the aging analysis, we sweep the PID replacement interval $LT_{pid}$ from 1 to 10 years to identify the carbon-optimal replacement interval per configuration. Aging rates are fixed as mentioned in Section~\ref{sec:aging_model}, deferring a sensitivity analysis of aging rates to future work.

\textbf{Manufacturing and Deployment Grid:}
Embodied carbon for the CPU, GPU, and other constituent components is obtained from prior works~\cite{gupta2022act,li2025ecoserve,thapamagar2026carbon,nvidia2025hgxh100pcf}. PID embodied carbon is evaluated under the NYUP grid (0.109 kgCO$_2$e/kWh), reflecting Micron's newly proposed fabrication facility in the region. For deployment, we consider Virginia's SRVC grid (0.269 kgCO$_2$e/kWh), which hosts the largest number of data centers in the United States. We defer a grid sensitivity analysis to future work.

\subsection{Comparison of Carbon Footprint Across PID Servers}

\begin{figure}[t]
    \centering
    \includegraphics[width=\linewidth]{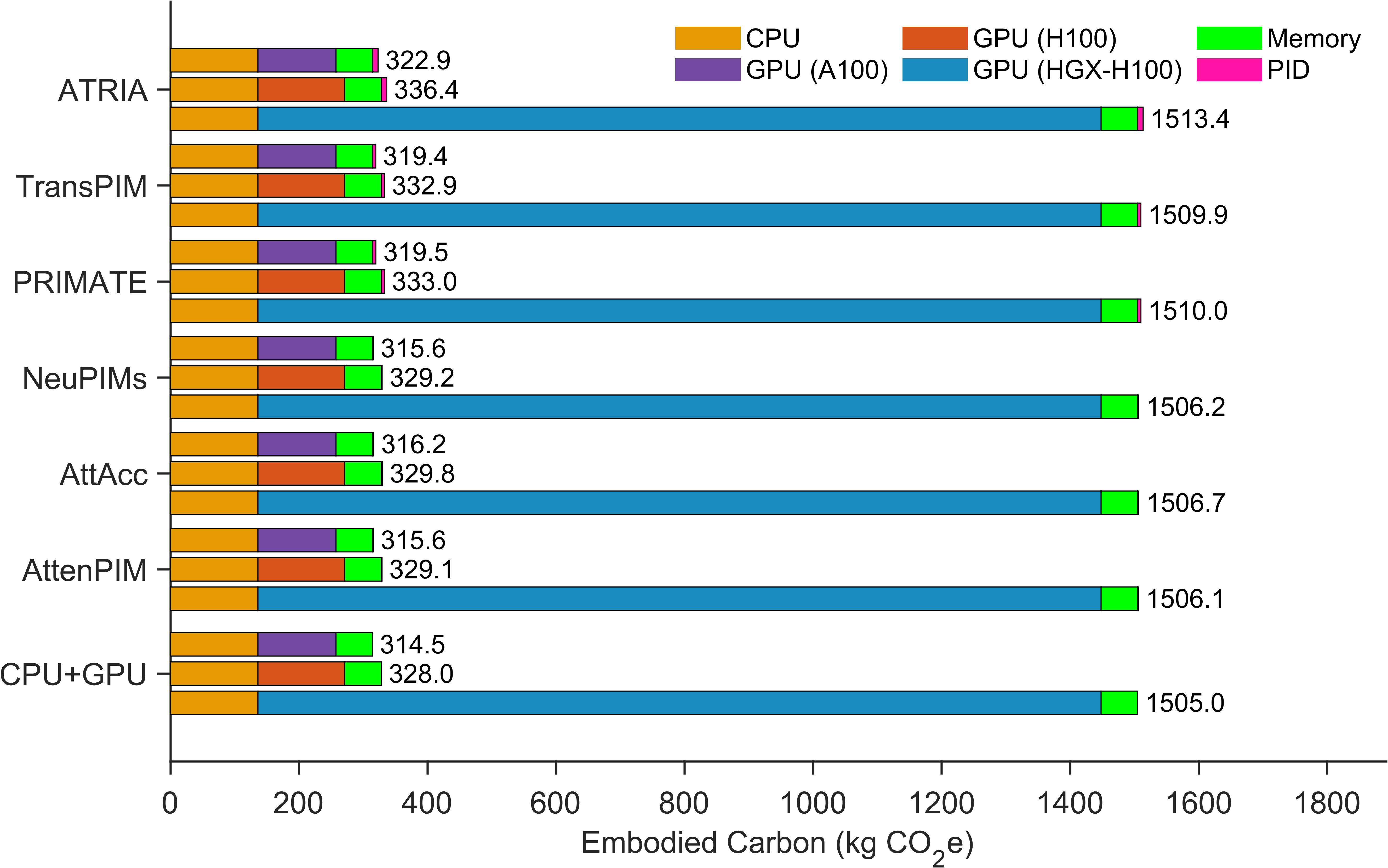}
    \caption{Embodied carbon of AI servers with different PID and GPU configurations. The host platform is the same, while PID and GPU selection affect the total embodied footprint.}
    \label{fig:embodied}
\end{figure}

\begin{figure}[t]
    \centering
    \subfloat[]{
        \includegraphics[width=\linewidth]{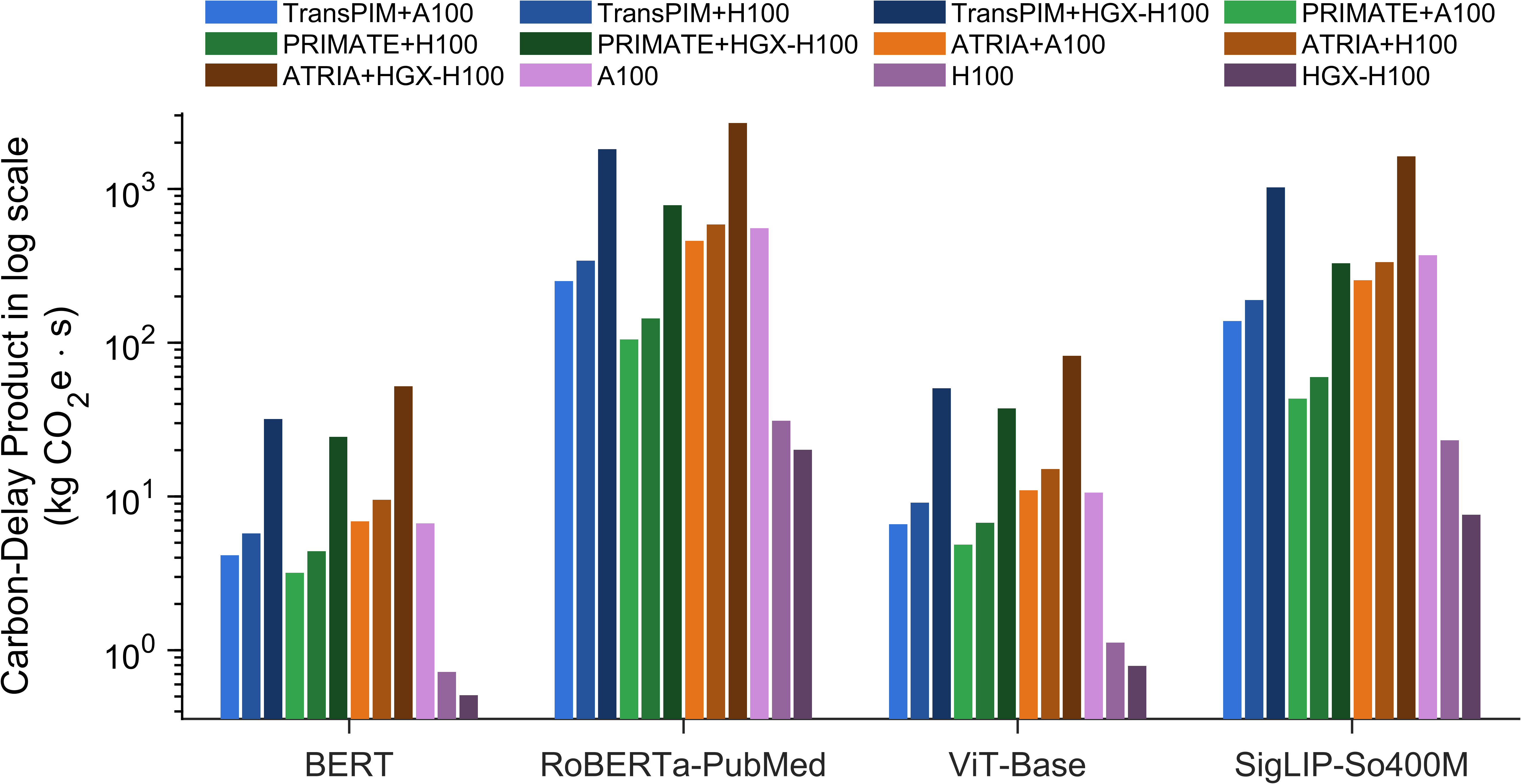}
        \label{fig:cdp_results_a}
    }
    \vspace{-4mm}
    \hfil
    \subfloat[]{
        \includegraphics[width=\linewidth]{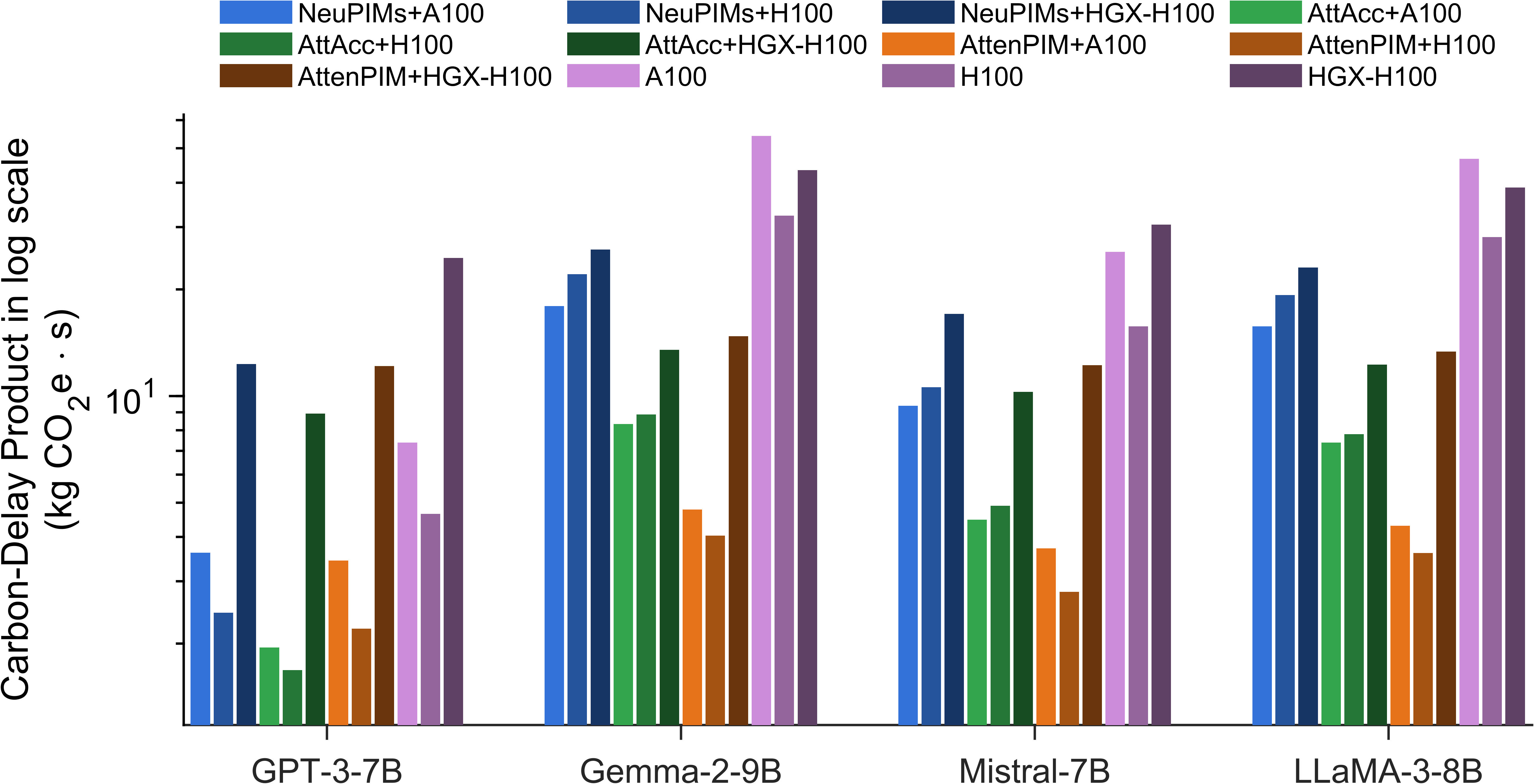}
        \label{fig:cdp_results_b}
    }
    \caption{Operational carbon-delay product for PID based servers. Panel (a) shows encoder workloads and panel (b) shows decoder workloads across all PID and GPU pairings.}
    \label{fig:cdp_results}
\end{figure}

\textbf{Embodied carbon:}  
Fig.~\ref{fig:embodied} shows that GPU choice dominates embodied carbon across PID–GPU combinations. The host system contributes a fixed 193.40~kgCO$_2$e in all configurations, while A100 has the lowest GPU embodied carbon, 1.1$\times$ below H100 and 10.8$\times$ below HGX-H100, whose eight H100 GPUs and NVLink and NVSwitch fabric drive its much larger footprint~\cite{nvidia2025hgxh100pcf}.

PID subsystems add only a small embodied overhead. AttenPIM has the lowest embodied carbon, 4.5$\times$, 4.7$\times$, 1.1$\times$, 1.6$\times$, and 7.9$\times$ lower than TransPIM, PRIMATE, NeuPIMs, AttAcc, and ATRIA, respectively, and even ATRIA raises total server embodied carbon by only about 2.5\%.

\textbf{Carbon-delay product:}  
Carbon-delay product is highly sensitive to the GPU baseline for encoder servers. Fig.~\ref{fig:cdp_results_a} shows that CDP rises with larger GPUs because they remain idle during PID-based prefill and idle power dominates. Averaged over workloads, moving from A100 to HGX-H100 increases CDP by 9.4$\times$ for TransPIM, 4.9$\times$ for PRIMATE, and 16.3$\times$ for ATRIA.

GPU-only encoder servers show the opposite trend. CDP falls as we move from A100 to HGX-H100 because higher GPU throughput shortens execution enough to offset higher power. Under A100, GPU-only has the highest CDP, exceeding ATRIA, TransPIM, and PRIMATE by 1.1$\times$, 1.9$\times$, and 3.8$\times$, respectively. These results suggest that pairing encoder PIDs with smaller, low idle power GPUs is more carbon efficient than using large GPUs that remain idle during prefill.

For decoder workloads in Fig.~\ref{fig:cdp_results}(b), GPU and PID share work sequentially, so faster execution by either reduces CDP. AttenPIM attains the lowest CDP under A100 and H100, outperforming NeuPIMs, AttAcc, and GPU-only by factors of 2.5 to 3.3, 1.2 to 1.6, and 5.2 to 6.5, respectively. Under HGX-H100, AttAcc edges out AttenPIM by about 1.2$\times$ while both still beat NeuPIMs and GPU-only by 1.4$\times$ to 2.6$\times$. This advantage reflects AttAcc's native one-to-one GPU to PIM pairing, which matches the multi-GPU structure of HGX-H100 and enables greater parallelism.

\begin{figure}[t]
    \centering
    \subfloat[]{
        \includegraphics[width=\linewidth]{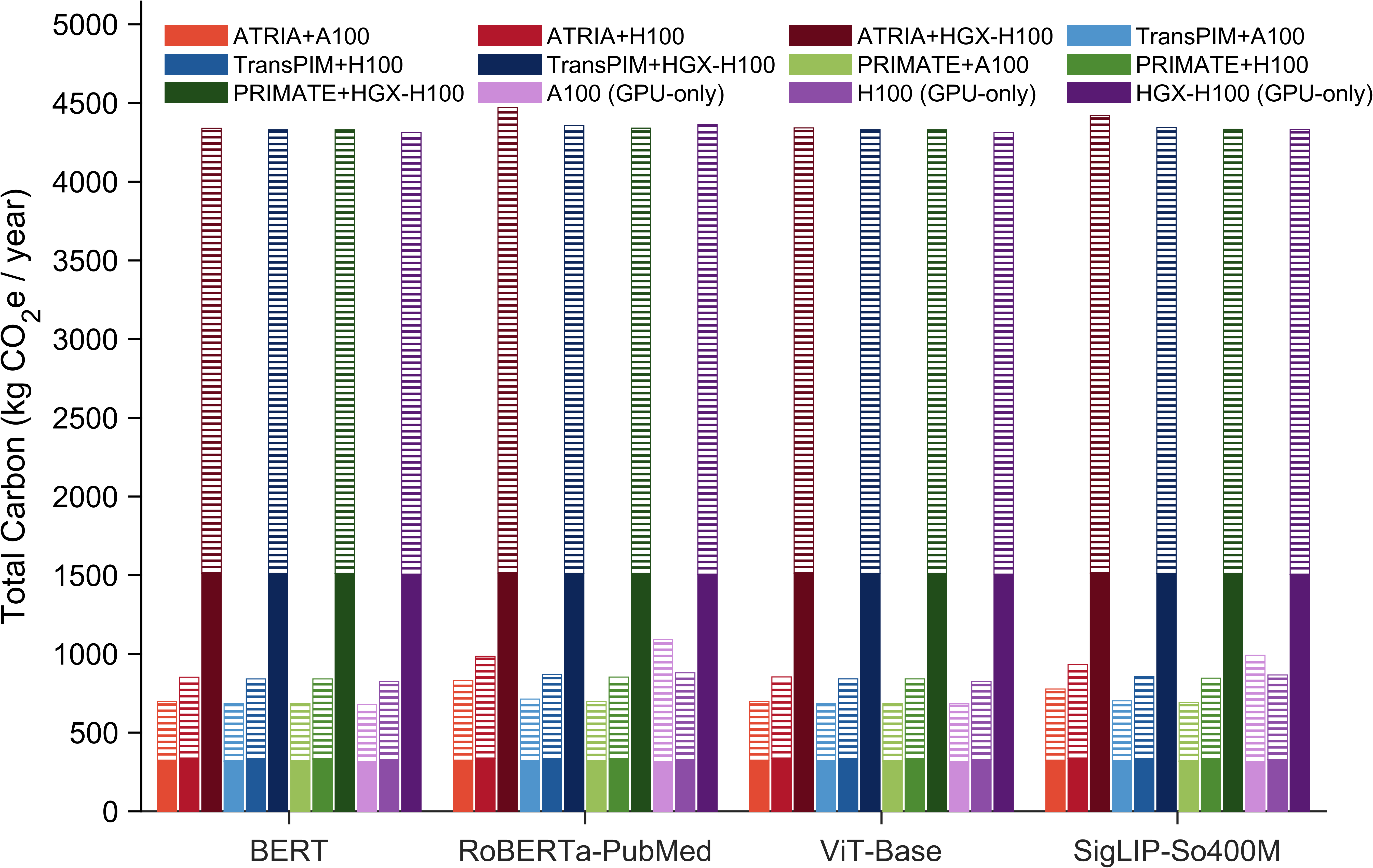}
        \label{fig:results_a}
    }
    \vspace{-4mm}
    \hfil
    \subfloat[]{
        \includegraphics[width=\linewidth]{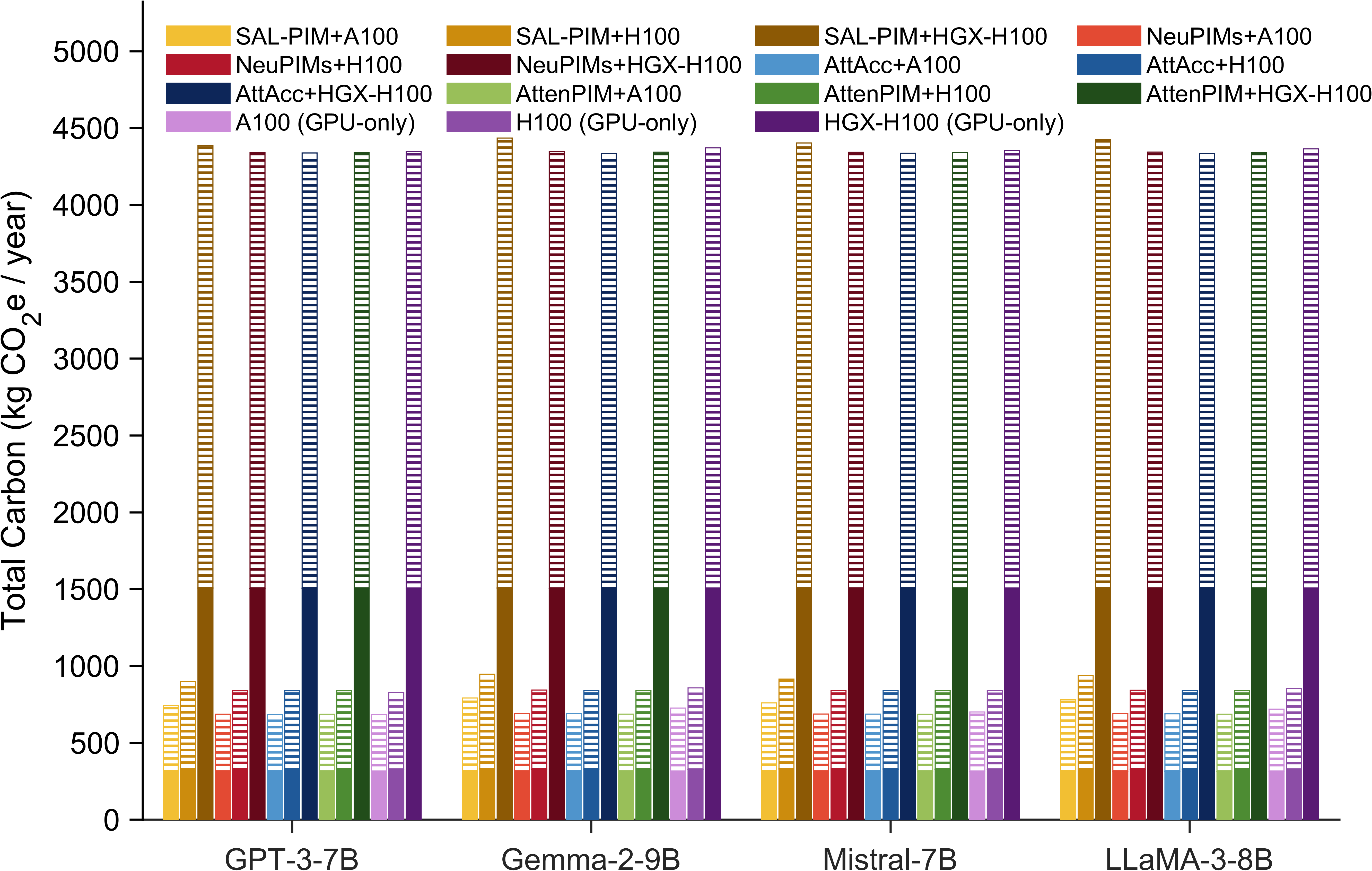}
        \label{fig:results_b}
    }
    \caption{Total carbon footprint of PID and GPU configurations for transformer inference. Panel (a) shows encoder servers and panel (b) shows decoder servers. Hatched bars on a white background indicate operational carbon and solid colored bars indicate embodied carbon.}
    \label{fig:results}
\end{figure}

\textbf{Total carbon:} Fig.~\ref{fig:results} combines embodied and operational terms into total carbon. Among GPU-only servers, HGX-H100 consistently has the highest total carbon, averaging 5.9$\times$ more emissions than A100 and 5.4$\times$ more than H100 across encoder and decoder workloads.  

For encoder servers, PRIMATE has the lowest total carbon on average, 1.01$\times$, 1.1$\times$, and 1.14$\times$ lower than TransPIM, ATRIA, and GPU-only, respectively. TransPIM and PRIMATE both maintain an advantage over GPU-only across all GPU baselines, although the margin shrinks for HGX-H100. ATRIA follows this trend at A100 and H100 but slightly exceeds GPU-only at HGX-H100, with a factor of 1.02$\times$.  

For decoder servers, AttenPIM has the lowest total carbon on average, 1.01$\times$, 1.01$\times$, and 1.04$\times$ below NeuPIMs, AttAcc, and GPU-only, respectively. NeuPIMs and AttAcc also have lower total carbon than GPU-only across all baselines, with their gains diminishing from A100 to HGX-H100.  

% For decoder servers, AttenPIM has the lowest total carbon on average, 1.01$\times$, 1.01$\times$, and 1.04$\times$ below NeuPIMs, AttAcc, and GPU-only, respectively. NeuPIMs and AttAcc also remain below GPU-only across all baselines, with their gains narrowing as the GPU scales from A100 to HGX-H100.  

Overall, PID integration consistently reduces total carbon relative to GPU-only baselines. This benefit, however, diminishes as the GPU grows larger. Under HGX-H100, the GPU's embodied and idle carbon dominate the footprint, and the differences between PID designs are very small. Hence, careful GPU selection is more important for sustainability than choosing any specific PID accelerator.

% Overall, PID integration consistently reduces total carbon relative to GPU-only at the A100 and H100 baselines, but the benefit diminishes as the GPU grows larger. Under HGX-H100, the host GPU's embodied and idle power dominate the footprint, so differences between PID designs nearly vanish and careful GPU selection becomes more important for sustainability than the specific choice of PID accelerator.

\subsection{Aging Analysis} \label{sec:aging_results}

\begin{figure}[t]
    \centering
    \subfloat[]{
        \includegraphics[width=\linewidth]{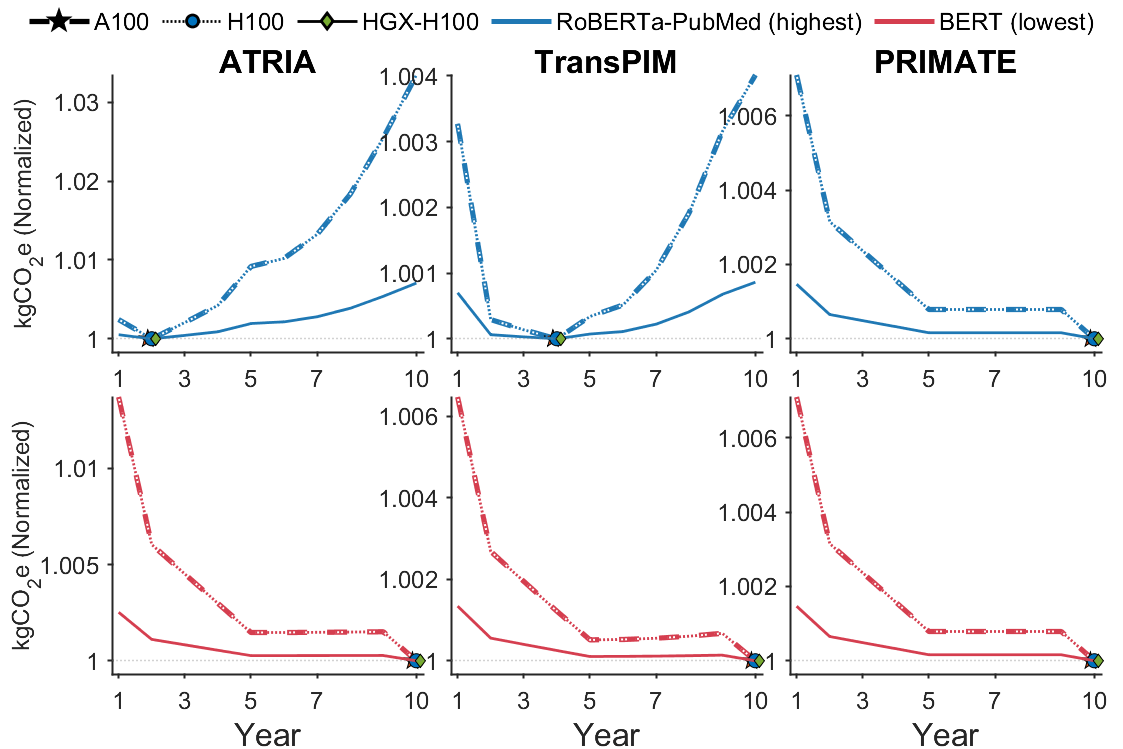}
        \label{fig:aging_a}
    }
    \vspace{-4mm}
    \hfil
    \subfloat[]{
        \includegraphics[width=\linewidth]{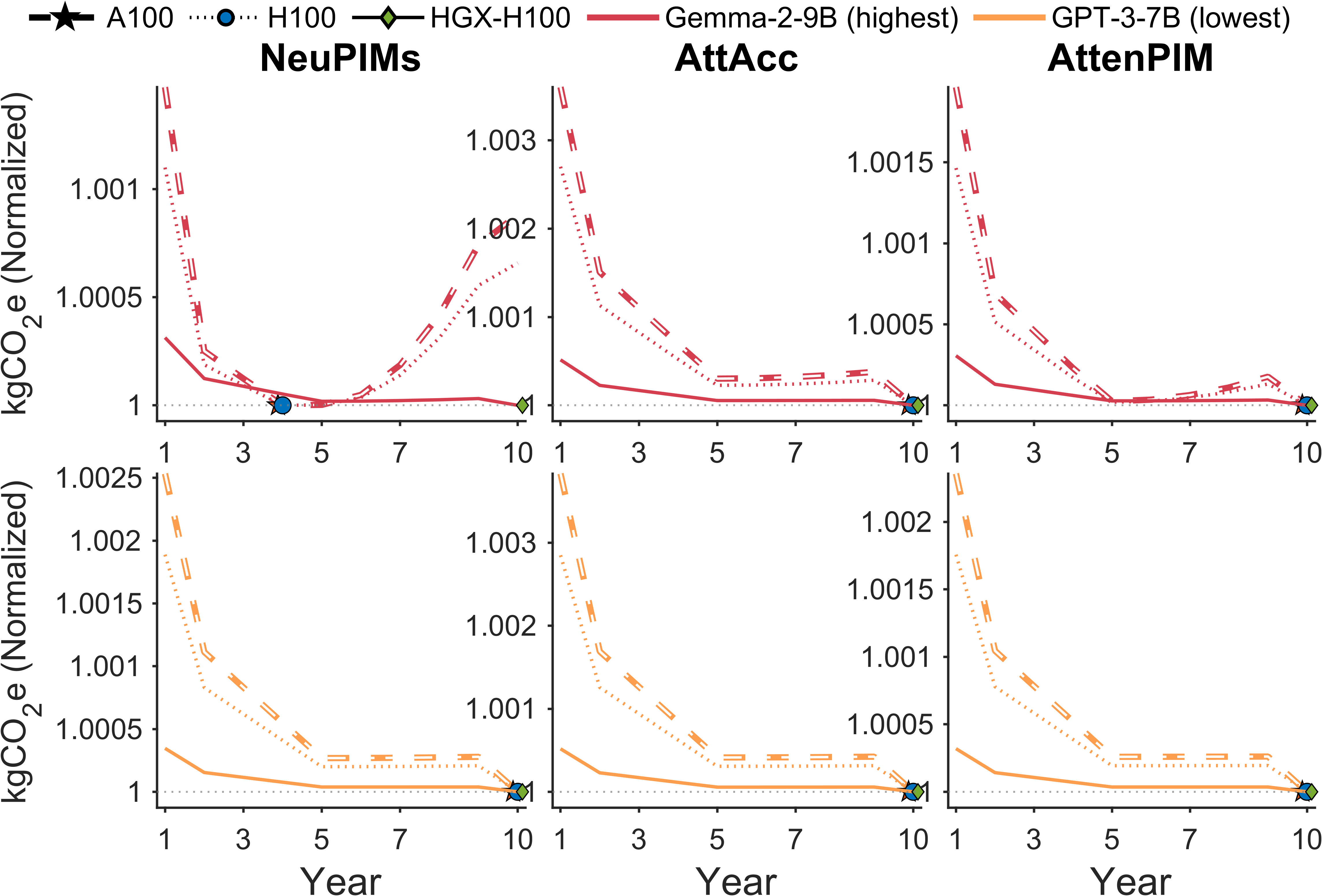}
        \label{fig:aging_b}
    }
    \caption{Impact of PID replacement interval on total carbon over a ten year deployment. Panel (a) shows encoder workloads and panel (b) shows decoder workloads. The top row in each panel uses the highest-carbon workload and the bottom row uses the lowest-carbon workload, with markers indicating the carbon-optimal replacement year for each GPU baseline.}
    \label{fig:aging}
\end{figure}

Fig.~\ref{fig:aging} examines total carbon as a function of PID replacement period $LT_{pid}$ for encoder and decoder workloads. For each PID, the top row uses the highest-carbon workload (RoBERTa-PubMed for encoder and Gemma-2-9B for decoder) and the bottom row uses the lowest-carbon workload (BERT for encoder and GPT-3-7B for decoder). The optimal replacement period differs across PID designs and GPU baselines. For encoder-oriented PIDs under RoBERTa-PubMed, the optimal $LT_{pid}$ is 2, 4, and 10 years for ATRIA, TransPIM, and PRIMATE, respectively. 
For the lower-carbon BERT workload, the optimal replacement period for all three designs is 10 years because reduced operational carbon incentivizes longer deployment periods. Across server configurations, the GPU choice has no impact on the optimal interval because GPUs are inactive in encoder servers and their total carbon contribution remains constant with $LT_{pid}$.
% Under the lower-carbon BERT workload, all three shift toward a 10-year interval because reduced operational carbon weakens the incentive to replace the PID subsystem. GPU baseline has no impact in this setting since the GPU is inactive for encoder servers and its carbon contribution does not change with $LT_{pid}$.  

Decoder-oriented PIDs show similar trends, with GPU effects most visible for NeuPIMs. Under Gemma-2-9B, the optimal replacement period for NeuPIMs is 4 years with A100 and H100 and 10 years with HGX-H100. AttAcc and AttenPIM have a 10-year replacement period across all GPU baselines because their operational carbon is low relative to the embodied overhead. GPT-3-7B exhibits a pattern similar to that of encoder servers, where all replacement periods move to 10 years due to the reduced operational carbon. Most replacement periods remain long because the PIDs in the decoder phase share work with GPUs. Since only the PIDs are replaced, replacement is preferred only when the PID's operational carbon is large relative to its embodied cost, as in NeuPIMs on Gemma-2-9B.  

% Decoder-oriented PIDs show similar trends, with GPU effects most visible for NeuPIMs. Under Gemma-2-9B, NeuPIMs prefers a 4-year replacement under A100 and H100 but a 10-year interval under HGX-H100. AttAcc and AttenPIM instead favor 10-year lifetimes across all GPU baselines, indicating that their operational carbon is low enough that frequent replacement offers little benefit. GPT-3-7B exhibits the same qualitative pattern. Splitting decoder work between GPU and PID reduces the PID's share of operational carbon, so replacement remains attractive only when the PID's operational footprint is large relative to its embodied cost, as with NeuPIMs on Gemma-2-9B.  

Across all designs, HGX-H100 servers have the longest replacement intervals. This is because their large embodied and idle contributions dominate the total carbon and diminish the benefit of frequent PID replacement. This highlights the need to choose GPU baselines carefully for inference servers with longer lifetimes.

% Across all designs, HGX-H100 favors the longest replacement intervals. Its large embodied and idle contributions dominate total carbon, reducing the benefit of frequent PID replacement and highlighting the need to choose GPU baselines carefully for long-lived inference servers.

\section{Conclusion}

This paper presented a carbon footprint analysis of PID-integrated transformer inference servers across three GPU baselines, six PID architectures, and eight workloads. PID subsystems add at most 2.5\% embodied overhead yet reduce total carbon by up to 1.14$\times$ compared with GPU-only configurations. GPU choice is the dominant factor for total carbon, with HGX-H100 systems incurring up to 4.9$\times$ higher CDP and 5.9$\times$ higher total carbon than A100 or H100 systems.

% This paper presented a carbon footprint analysis of PID-integrated transformer inference servers across three GPU baselines, six PID architectures, and eight encoder and decoder workloads. PID subsystems add at most 2.5\% embodied overhead yet reduce total carbon by up to 1.14$\times$ compared with GPU-only configurations using the same GPU. GPU choice is the dominant driver of absolute emissions, with HGX-H100 systems incurring up to 4.9$\times$ higher carbon-delay product and 5.9$\times$ higher total carbon than A100 or H100 systems.
As GPUs become more powerful, the carbon-delay product advantage from encoder-oriented PIDs diminishes, whereas decoder-oriented PIDs still retain up to 6.5$\times$ lower carbon-delay product than the GPU-only baseline under A100. Aging-aware analysis shows that the carbon-optimal PID replacement interval is shorter for PIDs with higher operational carbon, and longer for low-footprint designs such as PRIMATE. For decoder-oriented PIDs, the carbon-optimal replacement period is longer because the PIDs have lower operational carbon, as they split the work with GPUs. These patterns remain consistent across workloads and GPU generations. Sustainable PID deployment therefore requires a joint selection of PID architecture, GPU baseline, and replacement policy, where GPU choice remains the most important factor, followed by PID choice and lifetime tuning for carbon savings.

% Encoder-oriented PIDs lose much of their carbon-delay advantage as GPUs scale up, whereas decoder-oriented PIDs retain up to 6.5$\times$ lower carbon-delay product than GPU-only under A100. Aging-aware analysis shows that the carbon-optimal PID replacement interval depends on the balance between embodied and operational carbon. It is shorter when PID operational carbon is high, and longer for low-footprint designs such as PRIMATE or when decoder work is split between GPU and PID as in NeuPIMs, AttAcc, and AttenPIM. These patterns are consistent across workloads and GPU generations. Sustainable PID deployment for transformer inference therefore requires joint selection of PID architecture, GPU baseline, and replacement policy, with GPU choice as the primary lever and PID integration and lifetime tuning providing additional workload-dependent carbon savings.

\section*{Acknowledgments}
The authors used Claude (Anthropic) to assist in improving the clarity and consistency of the writing in selected sections of this manuscript. All technical content, results, figures, analyses, and conclusions are the sole work of the authors.

\bibliographystyle{IEEEtran}
\bibliography{references}

@techreport{chatterji2025chatgpt,
  author = {Chatterji, A. and others},
  title = {How People Use ChatGPT},
  institution = {National Bureau of Economic Research},
  number = {w34255},
  year = {2025}
}

@article{li2025ai,
  author = {P. Li and others},
  title = {Making AI less 'thirsty'},
  journal = {Commun. ACM},
  year = {2025},
  volume = {},
  number = {},
  pages = {}
}

@techreport{iea2024,
  author = {{International Energy Agency}},
  title = {Energy and AI: Executive Summary},
  institution = {IEA},
  year = {2024},
  type = {Analysis},
  address = {Paris, France},
  url = {https://www.iea.org/reports/energy-and-ai/executive-summary},
  note = {Accessed: 2026-04-25}
}

@article{shehabi2024,
  title = {2024 United States Data Center Energy Usage Report},
  author = {Shehabi, A. and others},
  year = {2024}
}

@article{liu2020energy,
  title = {Energy consumption and emission mitigation prediction based on data center traffic and PUE for global data centers},
  author = {Liu, Y. and others},
  journal = {Global Energy Interconnection},
  volume = {3},
  number = {3},
  pages = {272--282},
  year = {2020},
  publisher = {Elsevier}
}

@article{eeckhout2022first,
  title = {A first-order model to assess computer architecture sustainability},
  author = {Eeckhout, L.},
  journal = {IEEE CAL},
  volume = {21},
  number = {2},
  pages = {137--140},
  year = {2022},
  publisher = {IEEE}
}

@inproceedings{gupta2022act,
  title = {ACT: Designing sustainable computer systems with an architectural carbon modeling tool},
  author = {Gupta, U. and others},
  booktitle = {Proc. ISCA},
  pages = {784--799},
  year = {2022}
}

@article{ollivier2022sustainable,
  title = {Sustainable AI processing at the edge},
  author = {Ollivier, S. and others},
  journal = {IEEE Micro},
  volume = {43},
  number = {1},
  pages = {19--28},
  year = {2022},
  publisher = {IEEE}
}

@inproceedings{bardon2020dtco,
  title = {DTCO including sustainability: Power-performance-area-cost-environmental score (PPACE) analysis for logic technologies},
  author = {Bardon, M. G. and others},
  booktitle = {Proc. IEEE IEDM},
  pages = {41--4},
  year = {2020},
  organization = {IEEE}
}

@inproceedings{ji2024scarif,
  title = {SCARIF: Towards carbon modeling of cloud servers with accelerators},
  author = {Ji, S. and others},
  booktitle = {Proc. ISVLSI},
  pages = {496--501},
  year = {2024},
  organization = {IEEE}
}

@article{huang2019gpu,
  author = {Y. Huang and B. Guo and Y. Shen},
  title = {GPU energy consumption optimization with a global-based neural network method},
  journal = {IEEE Access},
  year = {2019},
  volume = {7},
  pages = {64303--64314}
}

@inproceedings{atria,
  author = {Shivanandamurthy, S. M. and others},
  title = {ATRIA: A bit-parallel stochastic arithmetic based accelerator for in-DRAM CNN processing},
  booktitle = {Proc. ISVLSI},
  year = {2021},
  pages = {200--205},
  doi = {10.1109/ISVLSI51109.2021.00045}
}

@article{star,
  author = {Afifi, S. and Thakkar, I. and Pasricha, S.},
  title = {STAR: A mixed analog stochastic in-DRAM convolutional neural network accelerator},
  journal = {IEEE MDAT},
  year = {2024},
  doi = {10.1109/MDAT.2024.3447580}
}

@article{artemis,
  author = {Afifi, S. and others},
  title = {ARTEMIS: A mixed analog-stochastic in-DRAM accelerator for transformer neural networks},
  journal = {IEEE TCAD},
  volume = {43},
  number = {11},
  pages = {3336--3347},
  year = {2024},
  doi = {10.1109/TCAD.2024.3446719}
}

@inproceedings{transpim,
  author = {M. Zhou and others},
  title = {TransPIM: A memory-based acceleration via software-hardware co-design for transformer},
  booktitle = {Proc. IEEE HPCA},
  year = {2022},
  pages = {1071--1085}
}

@inproceedings{primate,
  title = {PRIMATE: Processing in memory acceleration for dynamic token-pruning transformers},
  author = {Pan, Y. and others},
  booktitle = {Proc. ASP-DAC},
  pages = {557--563},
  year = {2024},
  organization = {IEEE}
}

@inproceedings{neupims,
  title = {NeuPIMs: NPU-PIM heterogeneous acceleration for batched LLM inferencing},
  author = {Heo, G. and others},
  booktitle = {Proc. ASPLOS},
  pages = {722--737},
  year = {2024}
}

@inproceedings{attenpim,
  title = {AttenPIM: Accelerating LLM attention with dual-mode GEMV in processing-in-memory},
  author = {Chen, L. and others},
  booktitle = {Proc. DAC},
  pages = {1--7},
  year = {2025},
  organization = {IEEE}
}

@inproceedings{attacc,
  title = {AttAcc! Unleashing the power of PIM for batched transformer-based generative model inference},
  author = {Park, J. and others},
  booktitle = {Proc. ASPLOS},
  pages = {103--119},
  year = {2024}
}

@article{li2025ecoserve,
  title = {EcoServe: Designing carbon-aware AI inference systems},
  author = {Li, Y. and others},
  journal = {arXiv preprint arXiv:2502.05043},
  year = {2025}
}

@article{ji2024towards,
  title = {Towards datacenter environmental sustainability using carbon depreciation models},
  author = {Ji, S. and others},
  journal = {arXiv preprint arXiv:2403.04976},
  year = {2024}
}

@inproceedings{masrur2012schedulability,
  title = {Schedulability analysis for processors with aging-aware autonomic frequency scaling},
  author = {Masrur, A. and others},
  booktitle = {Proc. IEEE RTCSA},
  pages = {11--20},
  year = {2012},
  organization = {IEEE}
}

@article{fieback2017dram,
  title = {DRAM reliability: Aging analysis and reliability prediction model},
  author = {Fieback, M.},
  journal = {Delft University of Technology},
  year = {2017}
}

@inproceedings{gupta2021chasing,
  title = {Chasing carbon: The elusive environmental footprint of computing},
  author = {Gupta, U. and others},
  booktitle = {Proc. IEEE HPCA},
  pages = {854--867},
  year = {2021},
  organization = {IEEE}
}

@inproceedings{zhao20243d,
  title = {3D-Carbon: An analytical carbon modeling tool for 3D and 2.5D integrated circuits},
  author = {Zhao, Y. and others},
  booktitle = {Proc. DAC},
  pages = {1--6},
  year = {2024}
}

@article{pope2023efficiently,
  title = {Efficiently scaling transformer inference},
  author = {Pope, R. and others},
  journal = {Proc. Mach. Learn. Syst.},
  volume = {5},
  pages = {606--624},
  year = {2023}
}

@article{park2022192,
  title = {A 192-Gb 12-high 896-GB/s HBM3 DRAM with a TSV auto-calibration scheme and machine-learning-based layout optimization},
  author = {Park, M.-J. and others},
  journal = {IEEE J. Solid-State Circuits},
  volume = {58},
  number = {1},
  pages = {256--269},
  year = {2022},
  publisher = {IEEE}
}

@article{magar2026dristi,
  title={DRISTI: Enabling Hierarchical Buffering and Flexible Dataflows for Reconfigurable In-DRAM Neural Network Accelerators},
  author={Magar, B.T. and others},
  journal={IEEE TCASAI},
  year={2026},
  publisher={IEEE}
}

@misc{samsung_lpddr5,
  author = {{Samsung Semiconductor}},
  title = {{LPDDR5} | {DRAM}},
  howpublished = {\url{https://semiconductor.samsung.com/dram/lpddr/lpddr5/}},
  year = {2026},
  note = {Product specifications and technical features. Accessed: 2026-04-26}
}

@misc{intel_xeon_max_9480,
  author = {{Intel Corporation}},
  title = {Intel® Xeon® {CPU} Max 9480 Processor (112.5M Cache, 1.90 GHz)},
  howpublished = {\url{https://www.intel.com/content/www/us/en/products/sku/232592/intel-xeon-cpu-max-9480-processor-112-5m-cache-1-90-ghz/specifications.html}},
  year = {2023},
  note = {Launched Q1 2023. Featuring 56 Cores and 64GB HBM2e. Accessed: 2026-04-26}
}

@article{nvidia_a100,
  author = {Choquette, J. and others},
  journal = {IEEE Micro},
  title = {NVIDIA A100 Tensor Core GPU: Performance and Innovation},
  year = {2021},
  volume = {41},
  number = {2},
  pages = {29--35},
  doi = {10.1109/MM.2021.3061394}
}

@mastersthesis{patel2025lifecycle,
  author  = {Patel, Samrat Pravin},
  title   = {Lifecycle Carbon Footprint and Sustainability Evaluation of {DRAM}-based Processing in Memory Computing Architectures},
  year    = {2025},
  school = {University of Kentucky},
  doi     = {10.13023/etd.2025.552},
  url     = {https://uknowledge.uky.edu/ece_etds/223},
  note    = {Theses and Dissertations--Electrical and Computer Engineering. 223}
}

@misc{nvidia2025hgxh100pcf,
  title        = {Product Carbon Footprint (PCF) Summary for NVIDIA HGX H100},
  author       = {{NVIDIA Corporation}},
  year         = {2025},
  howpublished = {\url{https://images.nvidia.com/aem-dam/Solutions/documents/HGX-H100-PCF-Summary.pdf}},
  note         = {Accessed: 2026-07-12}
}

@inproceedings{thapamagar2026carbon,
  title={Carbon at the memory wall: Lifecycle and aging-aware analysis of processing-in-{DRAM} {LLM} servers},
  author={Thapa Magar, B. and Patel, S. P. and Thakkar, I.},
  booktitle={Proc. IGSC},
  year={2026}
}

@techreport{nvidia2022h100whitepaper,
  title       = {NVIDIA H100 Tensor Core GPU Architecture},
  author      = {{NVIDIA Corporation}},
  year        = {2022},
  institution = {NVIDIA},
  note        = {Whitepaper},
  url         = {https://resources.nvidia.com/en-us-hopper-architecture/nvidia-h100-tensor-c}
}

@article{tu2024overview,
  title={An overview of large {AI} models and their applications},
  author={Tu, X. and others},
  journal={Visual Intelligence},
  year={2024},
  pages={34}
}

@inproceedings{sting,
  title={{STING}: A stochastic in-{DRAM} accelerator for graph neural networks},
  author={Afifi, S. and others},
  booktitle={Proc. GLSVLSI},
  year={2026},
  pages={398--404}
}

\vfill

\end{document}